\documentclass[journal,10pt,twoside]{IEEEtran}

\usepackage{amsmath, amssymb, amsfonts}
\usepackage{bm}
\usepackage{textcomp}
\usepackage{xcolor}
\usepackage{graphicx}
\usepackage{microtype}
\usepackage{booktabs}
\usepackage{cite}
\usepackage{algorithm}
\usepackage{algorithmic}
\usepackage{xcolor}
\usepackage{soul}
\usepackage{todonotes}
\usepackage{amsfonts}
\usepackage{multirow}
\usepackage[T1]{fontenc}

\usepackage[
  colorlinks=true,
  linkcolor=blue!60!black,
  citecolor=blue!60!black,
  urlcolor=blue!60!black
]{hyperref}

\begin{document}

\title{Histogramless Time-Domain Sketched\\
       Fluorescence Lifetime Imaging}

\author{Zhenya~Zang,~\IEEEmembership{}
        Istvan~Gyongy$^{\dagger}$,~\IEEEmembership{Senior Member,~IEEE,}
        ~and Mike~Davies$^{\dagger}$,~\IEEEmembership{Fellow,~IEEE}
\thanks{$^{\dagger}$These authors jointly supervised this work and 
        are co-senior authors.
        The authors are with the School of Engineering,
        University of Edinburgh, Edinburgh EH9 3FF, U.K.
        (e-mail: zzang@ed.ac.uk; igyongy2@exseed.ed.ac.uk; Mike.Davies@ed.ac.uk).} 
}


\maketitle

\begin{abstract}
We present a statistics-aware compression strategy that processes photon timestamps directly from time-correlated single-photon counting (TCSPC) modules for time-domain fluorescence lifetime imaging (FLIM). Rather than storing or transmitting the full histogram per pixel, timestamps are projected onto sparse, non-uniform one-dimensional spline sketches, with knot positions optimally allocated based on Fisher information. This knot allocation concentrates sketch channels where the decay signal exhibits the greatest statistical discriminability, rather than using a uniform allocation. The proposed approach is extensively validated on synthetic mono- and bi-exponential decay data and on experimental fluorescent dye data, demonstrating comparable accuracy to full-histogram non-linear least-squares fitting (NLSF) and Poisson maximum-likelihood estimation (MLE) at compression ratios of up to 256$\times$. We further validate the feasibility of integrating the timestamp-to-sketch projection directly into firmware via fixed-point (FXP) lookup-table (LUT) simulation, targeting high-spatial- resolution single-photon avalanche diode (SPAD) arrays subject to significant data-throughput constraints.
\end{abstract}

\begin{IEEEkeywords}
Fluorescence lifetime imaging microscopy (FLIM),
time-correlated single-photon counting (TCSPC),
Histogramless compression,
single-photon avalanche diode (SPAD).
\end{IEEEkeywords}

\section{Introduction}
\label{sec:intro}

Fluorescence lifetime imaging (FLIM) is a powerful label-free imaging modality that characterizes the molecular microenvironment of a fluorophore through the time constant of its excited-state decay, independent of fluorophore concentration \cite{lakowicz2006principles, torrado2024fluorescence}. Time-domain FLIM using time-correlated single-photon counting (TCSPC) offers high photon efficiency and temporal precision~\cite{becker2005advanced}, and is the modality to resolve multi-exponential decays arising from Förster Resonance Energy Transfer (FRET) \cite{sahoo2011forster} and metabolic activity quantification \cite{shirshin2022label}. Scanning-based FLIM systems using single-point detectors such as photomultiplier tubes (PMTs) offer a well-characterized instrument response function (IRF), but the throughput is fundamentally limited by the sequential acquisition process, meaning that imaging of large tissue samples takes hours per acquisition \cite{wang2024deep}. Scanning-free SPAD array-based FLIM systems address this by enabling parallel pixel acquisition, substantially reducing acquisition time~\cite{henderson2019192}. However, as the spatial resolution of the SPAD array increases, a new bottleneck emerges at the sensor-to-host data interface, where TCSPC produces a full histogram per pixel, typically comprising hundreds to thousands of time bins. Even at modest array sizes, SPAD sensors can generate tens of GB/s of raw histogram data~\cite{henderson2019192}, creating a fundamental throughput bottleneck for real-time lifetime reconstruction. This challenge will intensify as next-generation arrays scale to higher pixel counts and finer temporal resolution, increasing the demand for high-fidelity compression algorithms and compact hardware to accelerate timestamp processing, thereby accelerating downstream clinical decision-making. 

Device-level compression strategies have been proposed to alleviate the data-transfer bottleneck via on-chip histogramming~\cite{erdogan2019cmos}. However, on-chip histogram memory imposes high hardware costs and power consumption, rendering this approach unscalable to SPAD arrays with high spatial resolution when combined with per-pixel time-to-digital converters (TDCs). Compression algorithms have also been explored, including bin-merging~\cite{zang2023compact}, histogram-to-Phasor compression~\cite{heliot2021simple}, and histogramless deep-learning (DL) strategies~\cite{lin2024coupling, lin2024spiking}, which reduce temporal resolution or avoid histogram construction. However, data-driven DL approaches lack interpretability and introduce additional training overhead. This work addresses the data-throughput bottleneck by presenting a statistics-aware, histogramless compression strategy that processes photon timestamps directly, inspired by our previous work on sketched LiDAR \cite{spline2024}, which uses linear spline-basis compression and has been demonstrated on SPAD firmware on a field-programmable gate array (FPGA) \cite{zang2026fpga}. We propose a non-uniform, Fisher-information-guided knot allocation that concentrates sketch channels where the decay signal is most informative, and demonstrates its accuracy and hardware efficiency. This work aims to validate the feasibility of integrating the compression algorithm in the imaging system without modifying either the SPAD chip or the communication protocol. The sketch projection is designed for implementation on an FPGA running firmware that flexibly sends configuration and timing signals to a SPAD array or serves as a compact processor. This paper makes the following contributions:
\begin{itemize}
    \item We explain the conceptual connection between the phasor (Phasor) transform~\cite{digman2008phasor} and linear B-spline sketches, showing that the phasor is a special case of an $M=1$ Fourier-based sketch approximation. We introduce statistics-aware, non-uniform, hardware-friendly linear-spline sketches for FLIM (Sketched-FLIM) that reconstruct fluorescence lifetime parameters from mono- and bi-exponential decays without requiring histogram storage.
    \item We extensively validate the proposed framework on synthetic mono- and bi-exponential decay data and on experimental fluorescent dye data, demonstrating reconstruction fidelity comparable to full-histogram non-linear least-squares fitting (NLSF) and Poisson maximum-likelihood estimation (MLE) at a compression ratio of up to $256\times$.
    \item We discuss the feasibility of implementing the sketch projection directly in SPAD array firmware on an FPGA, eliminating the need for histogram memory. A fixed-point (FXP) LUT-based firmware simulation provides preliminary results across varying LUT depths, validating the proposed Sketched-FLIM for FPGA implementation.
\end{itemize}

\section{Problem Definition}
\subsection{Connections Between Fourier Sketch and Phasor}
\label{sec:model}
At a pixel, the ideal (unconvolved) single-exponential fluorescence decay is
\begin{equation}
    f(t) = e^{-(t-t_0)/\tau}, \qquad t \ge t_0,
    \label{eq:mono_decay}
\end{equation}
where $\tau$ is the lifetime, and $t_0$ is the
bin-index of the IRF peak. Let $I$ denote the IRF. The measured signal is the convolution
\begin{equation}
    h(t) = (I * f)(t) + b,
    \label{eq:convolution}
\end{equation}
with constant background $b$. In this work, we neglect $b$, as it can be largely suppressed by applying narrow-band emission filters in the optical path. Furthermore, in typical TCSPC-FLIM acquisitions, the fluorescence amplitude $A$ is sufficiently large relative to any residual background that the contribution of $b$ is negligible. Omitting this parameter reduces the degrees of freedom in the regression, leading to a more robust and stable lifetime estimation.
Assume that histograms from an SPAD array have $N$ time bins and bin width $\Delta t$, with bin centers $t_k = (k-\tfrac{1}{2})\Delta t$, $k=1,\dots,N$. Define the discrete IRF as $I_k = I(t_k)$ and the exponential basis as $e_k(\tau) = e^{-(t_k - t_0)/\tau}$.
The discretized convolution is
\begin{equation}
    g_k(\tau) = (I * e(\tau))_k
              = \sum_{j=1}^{N} I_j\, e_{k-j+1}(\tau).
\end{equation}
We normalize $g$ such that $\max_k g_k(\tau) = 1$, making $A$ the expected peak photon count. The expected photon count in bin $k$ is $\mu_k = A\, g_k(\tau)$. Under TCSPC acquisition, the histogram entries follow independent
Poisson statistics
\begin{equation}
    y_k \sim \mathrm{Poisson}(\mu_k), \qquad k=0,\dots,N-1,
    \label{eq:Possion_eq}
\end{equation}
and we denote the observed histogram by
$\bm{y}=(y_1,\dots,y_N)^\top$.
The histogram $\bm{y}$ is constructed from photon arrival times
$\{t_i\}_{i=1}^{P}$, where $P=\sum_k y_k$.

\subsection{Spline sketches as a generalization of Phasor}
Both the phasor method ~\cite{digman2008phasor} and the Fourier sketch~\cite{sheehan2021sketching} represent each per-pixel decay
as Fourier coefficients of the photon arrival time distribution.
The phasor computes a single complex coefficient (the fundamental Fourier coefficient) of the histogram $\bm{y} \in \mathbb{R}^N$ corresponding to the first harmonic of the histogram, 
\begin{equation}
    z = \sum_{k=0}^{N-1} y_k\, e^{-j2\pi k / N},
    \label{eq:phasor}
\end{equation}
where $y_k$ is the photon count in bin $k$. Each pixel is thereby
mapped to a single point $z$ in the phasor plot, from which the lifetime can be read directly from the semi-circle for mono-exponential decays, as explained in Appendix \ref{app:phasor_A}. For multi-component decays, a single complex number is insufficient, and additional geometric decomposition is required. The Fourier sketch addresses this limitation by randomly selecting $M$ of Fourier coefficients instead of one, evaluated at $M$ harmonics:
\begin{equation}
    s_m = \sum_{k=0}^{N-1} y_k\, e^{-j2\pi m k / N}.
    \qquad m = 1,\dots,M,
    \label{eq:fourier_sketch}
\end{equation}
Comparing \eqref{eq:phasor} and \eqref{eq:fourier_sketch}, it is
clear that $s_1 = z$: the phasor is the $M=1$ special case
of the Fourier sketch. Increasing $M$ captures higher-frequency
components of the decay, improving lifetime discrimination at the
cost of additional storage and computation. \ref{eq:fourier_sketch} is the direct multi-harmonic counterpart of the phasor sum. Increasing $M$ recovers progressively more information about the decay shape. Sketched-LiDAR \cite{spline2024} simplifies the Fourier sketch by replacing the complex exponential basis with linear, piecewise B-splines for depth estimation, where overlapping sketches capture relative statistical information about the timestamps of echo signals. For the linear B-spline, each basis function is a piecewise-linear triangle spanning two knot intervals. Like Fourier sketches, the B-spline sketch accumulates directly from raw timestamps, but requires only a single comparison and subtraction per photon per channel rather than computing trigonometric functions. The linear spline sketch is also an ideal candidate for convolving exponential decays; however, it requires knot placement informed by the distributional profile of decay signals and prior knowledge of the lifetime parameter range.

Applying the phasor method to bi-exponential decays,
\begin{align}
    f(t) &= \alpha_{1}\, e^{-(t - t_0)/\tau_1}
          + (1-\alpha_{1})\, e^{-(t - t_0)/\tau_2},
          \quad t \ge t_0,
    \label{eq:bi-exp_model}
\end{align}
where $\alpha_1 = 1 - \alpha_2$ is the fractional amplitude weighting the corresponding lifetime component, is challenging, as measured phasor points fall inside the universal semicircle at the intensity-weighted average of the two pure-component phasors. Although $\tau_1$, $\tau_2$, and $\alpha_1$ can be recovered geometrically via the lever rule~\cite{malacrida2021phasor}, the pure-component phasor positions from separate reference measurements are needed. The single Fourier harmonic capture only the first-order phase information, sufficient for mono-exponential decays but insufficient to resolve multi-component lifetime distributions. Multi-harmonic extensions recover more temporal information at the cost of additional channels
and loss of the intuitive geometric interpretation in the phasor plot.
Unlike Phasor, which is primarily designed for fast, fitting-free lifetime visualization, the proposed Sketched-FLIM aims to reconstruct lifetime parameters precisely, by encoding statistical information via spline basis functions $\phi_{i,p}(t)$ with only weak prior knowledge of the lifetime range, and recovers parameters via sketch-domain nonlinear least-squares optimization. Based on a $\phi_{i,p}(t)$ that will be detailed in the next section, $M$-dimensional feature vector for a single time $t$ is
\begin{equation}
    \bm{\Phi}_p(t) =
    \bigl[\phi_{0,p}(t),\;
           \phi_{1,p}(t),\;
           \dots,\;
           \phi_{M-1,p}(t)\bigr]^\top
    \in \mathbb{R}^M.
\end{equation}
Given photon arrival times $\{t_i\}_{i=1}^P$, we define the total spline sketch as
\begin{equation}
    \bm{s}
    = \sum_{i=1}^P \bm{\Phi}_p(t_i)
    \in \mathbb{R}^M.
    \label{eq:sketch_sum}
\end{equation}
For offline evaluation on synthetic histograms $\mathbf{g}(\tau)$, histograms can be converted to its sketch representation $\bm{s}(\tau)$ via the matrix $\bm{W}\in\mathbb{R}^{M\times N}$:
\begin{equation}
    \bm{s}(\tau) = \bm{W}\, \bm{g}(\tau),
    \label{eq:sketch_matrix}
\end{equation}
with entries $W_{i+1,k} = \phi_{i,p}(t_k), i=0,\dots,M-1, k=1,\dots,N.$ It is worth mentioning that the phasor compression is a projection of  $t_k$ that contributes $e^{j2\pi t_k / N}$ to the phasor sum, which can be accumulated on the fly without building the histogram. 

Sketched-LiDAR \cite{spline2024} used linearly allocated spline sketches, which would underperform in FLIM due to its exponential decay characteristics. In LiDAR, the spline sketch is designed to estimate the mean arrival time of a unimodal echo pulse. Uniformly spaced knots are effective because the echo can arrive anywhere within the measurement range with nearly equal prior probability~\cite{spline2024}. FLIM presents a fundamentally different estimation problem, in which the decay signal is a convolution of the IRF with a multi-exponential model that involves more parameters with non-uniform temporal distributions. The IRF peak occupies the earliest bins but carries little information about the lifetime; the discriminative information lies in the decay tail, meaning that uniform knots waste sketch channels on the IRF-dominated region while undersampling the tail. This mismatch motivates a non-uniform allocation of knots derived from intrinsic probability in convolved decays.

\subsection{Fisher Information--Based Knot Allocation}
\label{sec:fisher_knots}
Fisher information provides a natural criterion for knots allocation~\cite{rao1945information, cramer1999}, where the per-bin Fisher information for the lifetime parameter $\tau$ quantifies how much each time bin constrains $\tau$ given the Poisson photon statistics. Because FLIM experiments have prior knowledge of the plausible lifetime range, for instance, $[\tau_{\min}, \tau_{\max}]$ for mono-exponential models, and $[\tau_{1,\min}, \tau_{1,\max}], [\tau_{2,\min}, \tau_{2,\max}], [\alpha_{\min}, \alpha_{\max}]$ for the bi-exponential model, Fisher information can be averaged over the parameter space, achieving a Fisher information density $\mathcal{F}(t)$ along the temporal dimension. A single $\tau$ from the mono-exponential model is adopted for theoretical illustration hereafter for simplicity. If the log-likelihood $\ell(\tau)$ of the Poisson distribution is sharply curved near the true value, a small change in $\tau$ produces a large change in the predicted data, meaning the bin is highly informative. Conversely, a flat log-likelihood means the bin barely distinguishes nearby values of $\tau$. Given \ref{eq:Possion_eq}, the probability of observing $Y_k = y_k$ photons at bin $k$ is
\begin{equation}
    P(y_k \mid \tau)
    = \frac{\mu_k(\tau)^{y_k}\, e^{-\mu_k(\tau)}}{y_k!},
    \label{eq:poisson_pmf}
\end{equation}
taking the logarithm gives the per-bin log-likelihood
\begin{equation}
    \ell_k(\tau) = y_k \log \mu_k(\tau) - \mu_k(\tau) - \log(y_k!).
    \label{eq:poisson_ll}
\end{equation}
The sensitivity of the log-likelihood (the score function) to $\tau$ is calculated by its derivative 
\begin{align}
    \frac{\partial \ell_k}{\partial \tau}
    = \frac{\partial}{\partial \tau}
       \Bigl[y_k \log \mu_k(\tau) - \mu_k(\tau)\Bigr] \notag 
    &= \frac{y_k}{\mu_k}\frac{\partial \mu_k}{\partial \tau}
       - \frac{\partial \mu_k}{\partial \tau} \notag \\
    &= \frac{y_k - \mu_k}{\mu_k}
       \cdot \frac{\partial \mu_k}{\partial \tau}.
    \label{eq:score}
\end{align}
where $\partial \mu_k / \partial \tau = A\, \partial g_k / \partial \tau$
measures how sensitively the convolved decay rate responds to a change
in $\tau$, and where large gradients mean that the bin is
informative. While the true score function requires knowledge of the correct model, our information-guided knot allocation approximates this optimal statistic from the data by leveraging biological prior knowledge of lifetime ranges. The Fisher information for bin $k$ is defined as the expected square of \ref{eq:score} under repeated experiments,
\begin{equation}
    \mathcal{F}_k(\tau)
    = \mathrm{Var}\!\left[\frac{\partial \ell_k}{\partial \tau}\right]
    = \mathbb{E}\!\left[
        \left(\frac{\partial \ell_k}{\partial \tau}\right)^{\!2}
      \right].
    \label{eq:fim_def}
\end{equation}
Substituting \eqref{eq:score} into \eqref{eq:fim_def},
\begin{equation}
    \mathcal{F}_k(\tau)
    = \mathbb{E}\!\left[
        \left(\frac{\partial \ell_k}{\partial \tau}\right)^{\!2}
      \right]
    = \mathbb{E}\!\left[
        \left(\frac{y_k - \mu_k}{\mu_k}
        \cdot \frac{\partial \mu_k}{\partial \tau}\right)^{\!2}
      \right].
      \label{eq:fisher_expectation}
\end{equation}
Since $\partial \mu_k/\partial \tau$ is a constant with respect to observed
$y_k$, \ref{eq:fisher_expectation} becomes
\begin{equation}
    \mathcal{F}_k(\tau)
    = \left(\frac{\partial \mu_k}{\partial \tau}\right)^{\!2}
      \mathbb{E}\!\left[
        \left(\frac{y_k - \mu_k}{\mu_k}\right)^{\!2}
      \right].
\end{equation}
Using $\mathrm{Var}[y_k] = \mu_k$ under the Poisson model,
\begin{align}
    \mathbb{E}\!\left[
        \left(\frac{y_k - \mu_k}{\mu_k}\right)^{\!2}
    \right]
    = \frac{1}{\mu_k^2}\,\mathbb{E}\!\left[(y_k - \mu_k)^2\right]
    = \frac{\mathrm{Var}[y_k]}{\mu_k^2}
    &= \frac{1}{\mu_k},
    \label{eq:poisson_var}
\end{align}
achieving the per-bin Fisher information contribution,
\begin{equation}
    \mathcal{F}_k(\tau)
    = \frac{\bigl(\partial\mu_k/\partial\tau\bigr)^2}
           {\mu_k + \varepsilon},
    \label{eq:fim_bin}
\end{equation}
with a small regularizer $\varepsilon > 0$. Contributions are averaged over $n_{\mathrm{grid}}$ uniformly sampled lifetime parameters drawn from their biologically plausible ranges, where $n_{\mathrm{grid}}$ controls the granularity of the parameter sampling and is computed offline ($n_{\mathrm{grid}} = 500$ in this work). This obtains the Fisher information density
\begin{equation}
    \bar{\mathcal{F}}(t_k)
    = \frac{1}{n_{\mathrm{grid}}}
      \sum_{s=1}^{n_{\mathrm{grid}}}
      \mathcal{F}_k(\tau).
    \label{eq:fim_avg}
\end{equation}
For the bi-exponential model, the unknown parameter vector is
\begin{equation}
    (\theta=(\tau_1,\tau_2,\alpha_1)^\top). 
\end{equation}
In this case, the Fisher information at each time bin is a matrix rather than a scalar. 
The per-bin Fisher information matrix is defined as
\begin{equation}
\mathcal{I}_k(\theta)
=
\frac{1}{\mu_k(\theta)+\epsilon}
\nabla_{\theta}\mu_k(\theta)
\nabla_{\theta}\mu_k(\theta)^\top
\end{equation}
To obtain a scalar information density for knot allocation, we use the trace of this matrix
\begin{equation}
\mathcal{F}_k(\theta)
=
\mathrm{tr}\{\mathcal{I}_k(\theta)\}
=
\frac{
\left(\frac{\partial \mu_k}{\partial \tau_1}\right)^2
+
\left(\frac{\partial \mu_k}{\partial \tau_2}\right)^2
+
\left(\frac{\partial \mu_k}{\partial \alpha_1}\right)^2
}
{\mu_k(\theta)+\epsilon}
\end{equation}
This trace criterion measures the total sensitivity of each time bin to all three bi-exponential parameters.

The Fisher information density is normalized to a probability distribution, and its cumulative distribution function (CDF) is formed as
\begin{equation}
    C(t_k)
    = \frac{\displaystyle\sum_{j=1}^{k}
            {\bar{\mathcal{F}}}(t_j)\,\Delta t}
           {\displaystyle\sum_{j=1}^{N}
            {\bar{\mathcal{F}}}(t_j)\,\Delta t}.
    \label{eq:fim_cdf}
\end{equation}
The $M+2$ knot boundaries $\{\xi_0,\xi_1,\ldots,\xi_{M+1}\}$ are then
placed at equally spaced quantile levels of $C$,
\begin{equation}
\xi_m = C^{-1}\left(\frac{m}{M+1}\right), \quad m = 0,1,\ldots,M+1.
    \label{eq:knot_placement}
\end{equation}
with $\xi_0 = t_1$ and $\xi_{M+1} = t_N$. By construction, each resulting spline channel integrates an equal amount of Fisher information, ensuring that every sketch dimension
contributes equally to lifetime estimation. We use linear piecewise B-splines, the $i$-th basis function is a piecewise-linear triangle defined by three consecutive knots
$\xi_i < \xi_{i+1} < \xi_{i+2}$,
\begin{equation}
    \phi_{i,1}(t) =
    \begin{cases}
        \dfrac{t - \xi_i}{\xi_{i+1} - \xi_i},
            & \xi_i \leq t < \xi_{i+1}, \\[8pt]
        \dfrac{\xi_{i+2} - t}{\xi_{i+2} - \xi_{i+1}},
            & \xi_{i+1} \leq t < \xi_{i+2}, \\[8pt]
        0, & \text{otherwise},
    \end{cases}
    \label{eq:linear_spline}
\end{equation}
where $i = 0,\ldots, M-1$, and the input t is clamped to the acquisition boundaries $[\xi_0,\xi_{M+1}]$. The knot boundaries $\{\xi_0, \xi_1, \dots, \xi_M\}$ are placed according to the Fisher-optimal allocation in \ref{eq:knot_placement}.
\begin{figure}[!t]
    \centering
    \includegraphics[width=0.98\linewidth]{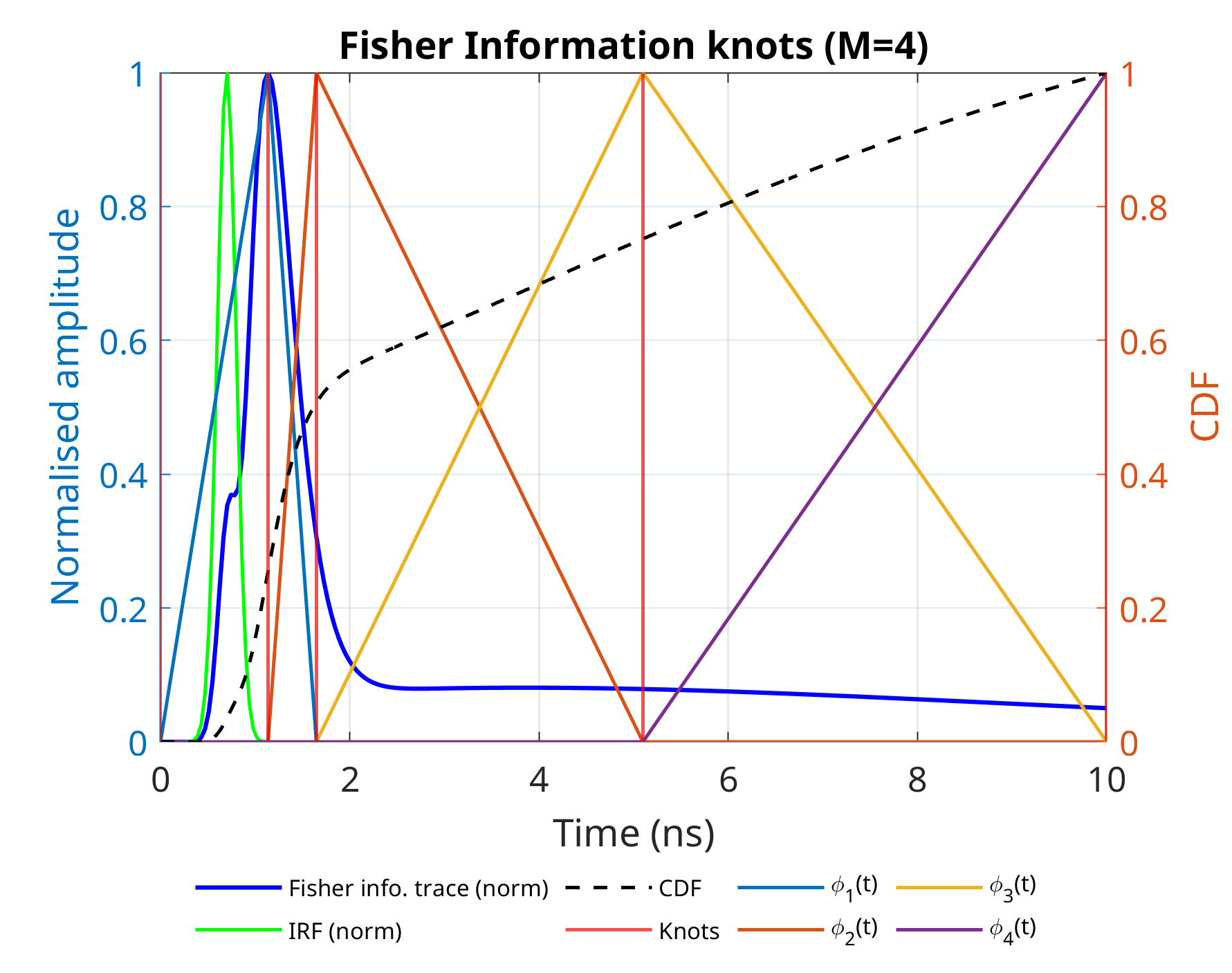}
    \caption{
    Fisher information-based knot placement for $M = 4$ sketch features.
    The normalized Fisher information density trace (blue) and IRF (green); the CDF. Knot boundaries (red vertical lines) are placed at equally spaced quantiles of the CDF; each interval between knots integrates an equal amount of Fisher information, and the $M$ overlapping triangle bases are constructed on these intervals.
    }
    \label{fig:Fisher_knots_example}
\end{figure}

\begin{figure}[!t]
    \centering
    \includegraphics[width=1\linewidth]{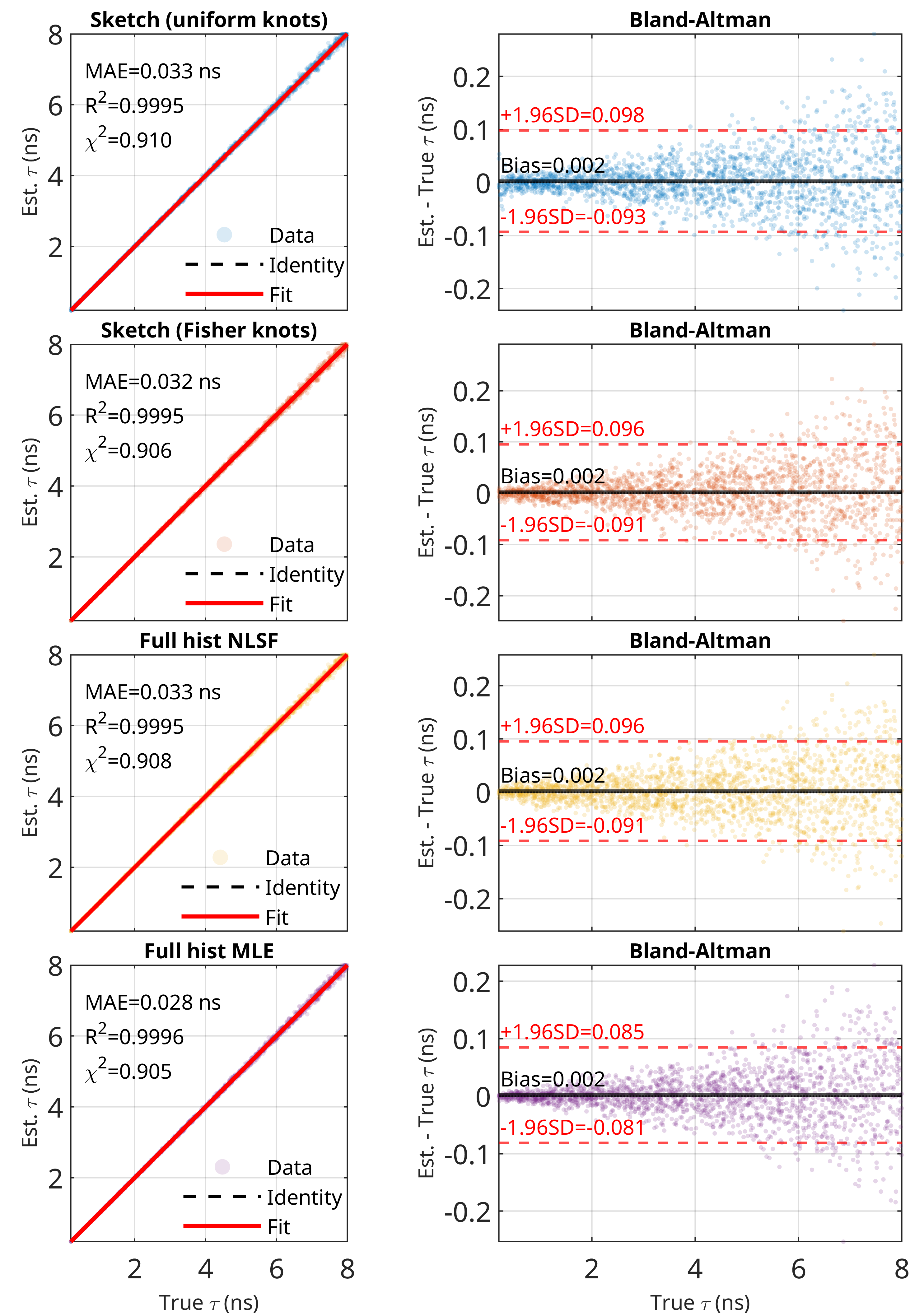}
    \caption{
    Lifetime reconstruction for mono-exponential decay across 2,000 synthetic samples ($A = 500$ photons, $M = 4$, $\tau \in [0.2, 8]$~ns).
    Left: estimated vs.\ true $\tau$ with identity (black dashed) and
    regression fit (red solid); MAE and $R^2$ annotated per panel.
    Right: Bland--Altman error plots with mean bias (black solid) and
    $\pm 1.96\,\mathrm{SD}$ limits (red dashed).
    Rows (1-4): uniform-knot sketch,
    Fisher-knot sketch, full-histogram NLSF, and full-histogram MLE. All metrics are computed over the 2,000 samples.
    }
    \label{fig:mono_results}
\end{figure}

\begin{figure*}[!t]
    \centering
    \includegraphics[width=1\textwidth]{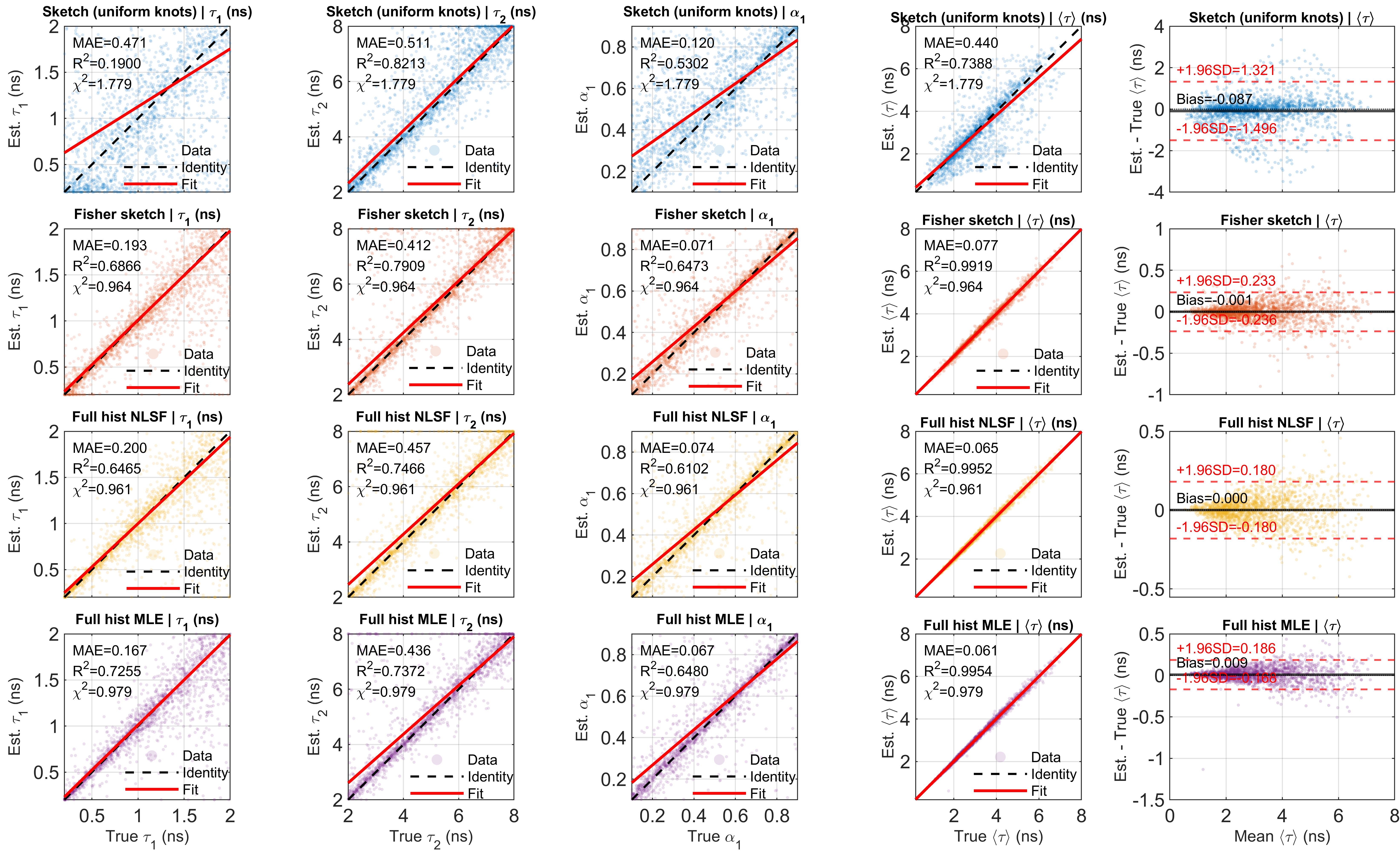}
    \caption{
    Lifetime retrieval for bi-exponential decays across 2,000 synthetic
    samples ($A = 500$, $M = 4$, $\tau_1 \in [0.2, 2]$~ns,
    $\tau_2 \in [2, 8]$~ns, $\alpha_1 \in [0.05, 0.95]$).
    Columns 1--4: estimated vs.\ true values for $\tau_1$, $\tau_2$,
    $\alpha_1$, and mean lifetime
    $\langle\tau\rangle = \alpha_1\tau_1 + (1-\alpha_1)\tau_2$,
    with identity (black dashed) and regression fit (red solid);
    MAE and $R^2$ annotated per panel.
    Column 5: Bland--Altman error of $\langle\tau\rangle$ with mean bias
    (black solid) and $\pm 1.96\,\mathrm{SD}$ limits (red dashed).
    Rows 1--4: uniform-knot sketch,
    Fisher-knot sketch, full-histogram NLSF, and full-histogram MLE. All metrics are computed over the 2,000 samples.}
    \label{fig:Bi_fitting_results}
\end{figure*}

\section{Lifetime Retrieval from the Sketch}
\label{sec:retrieval}
This section introduces lifetime reconstruction using statistics-aware linear knot allocation for both mono- and bi-exponential decays. As bi-exponential decays can well approximate signals arising from multi-exponential decays \cite{li2020investigations}, only these two models are considered in this study.
\subsection{Mono-Exponential Model}
Let $\tilde{\mathbf{s}}_{\mathrm{meas}}\in\mathbb{R}^M$ be the
normalized sketch computed from the measured timestamps. Based on \ref{eq:sketch_matrix}, we define the squared error of the sketch-domain (SSE)
\begin{equation}
    \mathrm{SSE}(\tau)
    = \bigl\|\tilde{\mathbf{s}}_{\mathrm{meas}}
             - \mathbf{s}(\tau)\bigr\|_2^2
    = \sum_{i=1}^M
      \bigl(\tilde{s}_{\mathrm{meas},i} - s_i(\tau)\bigr)^2.
    \label{eq:sse_def_all}
\end{equation}
Treating $\tau$ as a continuous parameter in $[\tau_{\min},\tau_{\max}]$, the lifetime is recovered by
\begin{equation}
    \tau_{\mathrm{fit}}^{\star}
    = \underset{\tau \in [\tau_{\min},\tau_{\max}]}{\arg\min}\;
      \mathrm{SSE}(\tau).
    \label{eq:tau_fit_only}
\end{equation}
This one-dimensional optimization is solved with
a trust-region-reflective nonlinear least-squares fitting (NLSF).

\subsection{Bi-Exponential Model}
\label{sec:Bi_Exp_parame}
 Unlike the mono-exponential model, the bi-exponential model retrieves two lifetimes and $\alpha_1$, thereby increasing the parameter search complexity, with the same NLSF. The optimization is initialized at the midpoint of the possible parameter range. Fig.~\ref{fig:Fisher_knots_example} illustrates the Fisher-based knot allocation when $M = 4$ for bi-exponential decays, with a constant IRF's full width at half maximum (FWHM) of 100~ps, peak photon counts $A=$200, and $N=256$. The per-bin Fisher information density $\mathcal{F}(t_k) \in \mathbb{R}^{N}$, computed via \ref{eq:fim_avg} and averaged over 500 randomly sampled parameter combinations $(\tau_1, \tau_2, \alpha_1)$. A peak occurs just after the IRF region ($\sim$1~ns), where photon arrival times carry the most discriminative information about $\tau$, followed by a gradual decay through the exponential tail. Normalizing $\mathcal{F}(t)$ 
to a probability density and forming its CDF $C(t)$ maps time to the 
cumulative fraction of total Fisher information. Placing knots at equally spaced quantile levels of $C(t)$ with $M$=4 shows three interior knots within the first 2~ns and one near 5~ns, concentrating sketch resolution where the lifetime discrimination is greatest. By contrast, uniform allocation spaces all boundaries evenly across the 10~ns window, wasting sketch dimensions on the low-information tail and undersampling the high-information early region.

\section{Synthetic Datasets Evaluation}
\subsection{Accuracy Evaluation}
\label{sec:synth_results}
Synthetic mono- and bi-exponential datasets are generated following
the IRF and noise models in \cite{zang2022fast}. Reconstruction accuracy is evaluated over 2,000 random trials for
each model using $M = 4$ sketch features, with full-histogram NLSF
and maximum likelihood estimation (MLE) as reference methods. For full-histogram NLSF, rather than simply comparing generated normalized ($A=1$) decays with the normalized ($A=1$) analytical model, $A$ can be optimized during each iteration using closed-form expressions. Suppose we minimize $\sum_{k=1}^{N} (y_k - A \cdot g_k)^2$ over $\tau_1, \tau_2, \alpha_1$. $A$ is solved by $\frac{\partial}{\partial A} \sum_{k} (y_k - A \cdot g_k)^2 = 0$, obtaining $A^* = \frac{y^T g}{g^T g}$. 
\begin{figure*}[!t]
    \centering
    \includegraphics[width=\textwidth]{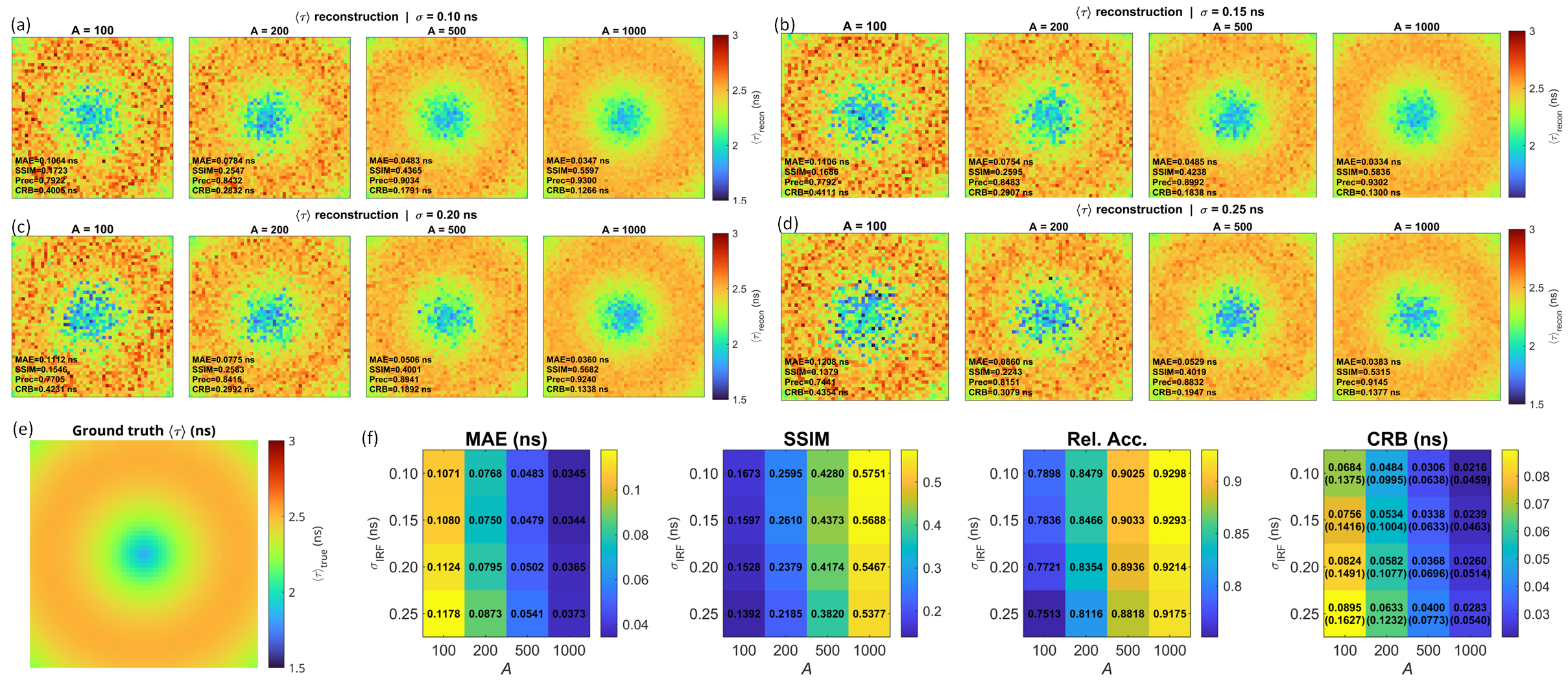}
    \caption{Reconstruction of mean lifetime $\langle\tau\rangle$
        from bi-exponential FLIM data across photon counts $A$ and
        IRF widths $\sigma_{IRF}$. $M=4$. (a)--(d)~Reconstructed $\langle\tau\rangle$ maps for $\sigma_{IRF} = 0.10$, $0.15$, $0.20$, and $0.25$\,ns,
        respectively, with MAE, SSIM, and relative accuracy.
        (e)~Ground-truth (GT) $\langle\tau\rangle$ map.
        (f)~Summary heatmaps of MAE\,(ns), SSIM, relative accuracy (1-RMSE of $\langle\tau\rangle$) over the range of $\langle\tau\rangle$, and CRB (ns) with RMSE in each configuration (in parentheses), as functions of $A$ and $\sigma_{IRF}$.}
    \label{fig:bi_A_sigma}
\end{figure*}

\begin{figure}[!t]
    \centering
    \includegraphics[width=1\linewidth]{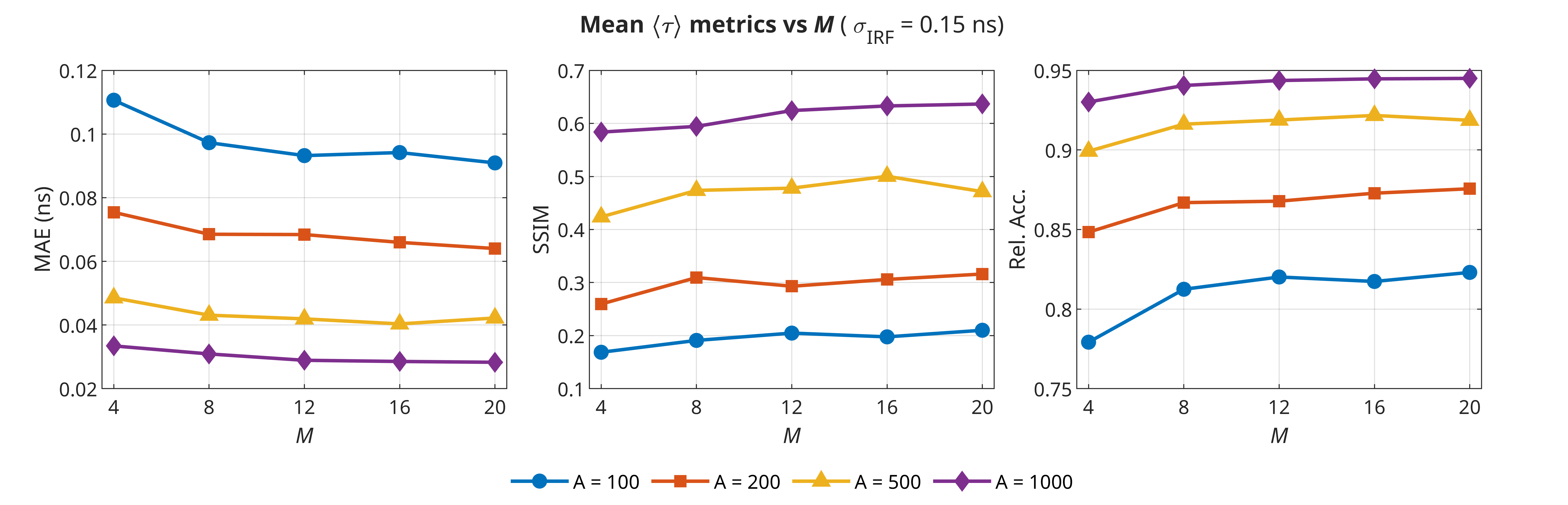}
    \caption{Mean lifetime $\langle\tau\rangle$ reconstruction metrics as a function of sketch dimension $M$, with one representative IRF $\sigma_{IRF}=0.15$ ns, for peak photon counts $A \in \{100, 200, 500, 1,000\}$.}
    \label{fig:mean_tau_vs_M}
\end{figure}

\begin{figure}[!t]
    \centering
    \includegraphics[width=1\linewidth]{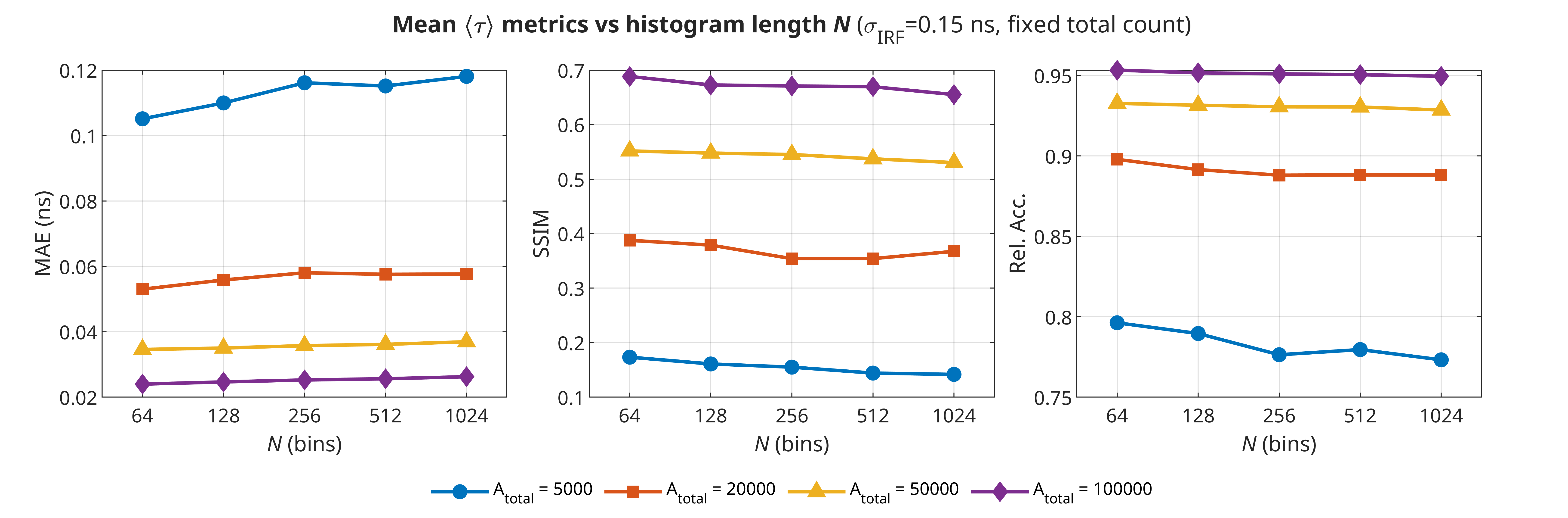}
    \caption{Mean lifetime $\langle\tau\rangle$ reconstruction metrics as a function 
    of histogram bin count $N$, with one representative IRF $\sigma_{IRF}=0.15$ ns, for fixed total photon counts $A_\text{total} \in \{5,000, 20,000, 50,000, 100,000\}$, with sketch dimension $M=8$.}
    \label{fig:hist_length}
\end{figure}

The same decay generation parameters are used as in \ref{sec:Bi_Exp_parame}. As shown in Fig.~\ref{fig:mono_results}, all four methods perform comparably for the mono-exponential model, with both sketch variants achieving accuracy close to full-histogram MLE (MAE $\approx 0.03$~ns, $R^2 > 0.999$), and the choice of knot placement has a negligible effect at this level of model simplicity. $\chi^{2} = \frac{1}{N} \sum_{k} \frac{(y_k - g_k)^2}{y_k}$ is used to evaluate the goodness of fitting decay. When forced to match both shape and area simultaneously with $A=1$, the NLSF optimizer underestimates $\tau$ to reduce tail mismatch at the cost of shape accuracy. $A^*$ compensates for area differences and allows the optimizer to focus purely on decay kinetics while providing unbiased estimates. The distinction becomes noticeable for the bi-exponential model as shown in Fig.~\ref{fig:Bi_fitting_results} evaluated with the same number of samples, photon counts, and $M$, with lifetimes parameter ranges $\tau_1 \in [0.2, 2]$~ns, $\tau_2 \in [2, 8]$~ns, $\alpha_1 \in [0.05, 0.95]$. Columns 1--4 show estimated vs.\ true values for $\tau_1$, $\tau_2$, $\alpha_1$, and the amplitude-weighted mean lifetime $\langle\tau\rangle = \alpha_1\tau_1 + (1-\alpha_1)\tau_2$. Column 5 shows Bland--Altman error plots for $\langle\tau\rangle$. The retrieval of $\tau_1$ is challenging for uniform knots ($R^2=0.19$, MAE $=0.47$), and $\alpha_1$ is challenging across all methods with $R^2$ values of $0.53$--$0.65$. Retrieval of $\tau_2$ is more reliable across all methods ($R^2 > 0.74$), as the long-lifetime component induces the decay tail, where the short component has already decayed, providing a cleaner signal for $\tau_2$ estimation. The mean lifetime $\langle\tau\rangle$ is the most robustly estimated, with $R^2 > 0.99$ for all methods except the uniform-knot sketch ($R^2 = 0.74$). Fisher-optimal knots reduce $\langle\tau\rangle$ MAE by $5.7\times$ over uniform knots ($0.077$ vs.\ $0.44$~ns), approaching full-histogram NLSF ($0.065$~ns) and MLE ($0.061$~ns) accuracy. The Bland--Altman plots confirm that the Fisher-sketch achieves near-zero bias ($-0.001$~ns) and tight limits of agreement ($[-0.24, +0.23]$~ns), close to NLSF ($\pm 0.18$~ns) and MLE ($[-0.17, +0.19]$~ns), whereas the uniform-knot sketch exhibits a substantial negative bias ($-0.087$~ns) and wide limits of agreement ($[-1.5,+1.3]$~ns). Although the Fisher-informed sketch achieves higher accuracy than full-histogram NLSF for individual lifetime parameters across the three metrics, because the sketch projection suppresses low-information, low-count bins that receive equal weight in the unweighted NLSF objective, full-histogram NLSF provides better \(\langle\tau\rangle\) estimation. This suggests that parameter errors in NLSF are compensated when computing \(\langle\tau\rangle\), whereas sketch-based methods achieve more independent parameter errors.

We also explore the performance of the maximum $\mathcal{F}_k(\tau)$ in \ref{eq:fim_avg} compared to the averaged version. The averaged version weights all sampled lifetime parameters equally, whereas the max version selects the highest Fisher information at each time point. Table \ref{tab:max_ave_Fisher} compares the performance between maximum and averaged $\mathcal{F}_k(\tau)$, using the same datasets in Fig. \ref{fig:Bi_fitting_results}. The table shows that averaging $\mathcal{F}_k(\tau)$ yields lower errors across all parameters, indicating more effective knots' allocation.

\begin{table}[t]
\centering
\caption{Comparison of Average vs Max $\mathcal{F}_k(\tau)$ ($M$=4).}
\label{tab:max_ave_Fisher}
\begin{tabular}{@{}lcccc@{}}
\toprule
\multirow{2}{*}{\textbf{Parameter}} & \multicolumn{2}{c}{\textbf{Max $\mathcal{F}_k(\tau)$}} & \multicolumn{2}{c}{\textbf{Average $\mathcal{F}_k(\tau)$}} \\
\cmidrule(lr){2-3} \cmidrule(lr){4-5}
& MAE & RMSE & MAE & RMSE \\
\midrule
$\tau_1$ (ns)        & 0.2467 & 0.4114  & \textbf{0.1881} & \textbf{0.3308}  \\
$\tau_2$ (ns)        & 0.6395 & 1.2160  & \textbf{0.3840} & \textbf{0.7735}  \\
$\alpha_1$           & 0.0917 & 0.1700  & \textbf{0.0650} & \textbf{0.1315}  \\
$\langle\tau\rangle$ (ns) & 0.0837 & 0.1169  & \textbf{0.0575} & \textbf{0.1165} \\
\bottomrule
\end{tabular}
\end{table}

\subsection{Evaluation of Varying Configurations}
\label{sec:photon_efficiency}
Sketched-FLIM is further evaluated across varying photon counts and IRF FWHMs ($\sigma_{IRF}$) to investigate its performance. As demonstrated in the previous section, the results of the Fisher-based sketch for mono-exponential decay processing differ marginally from uniform knots and full-histogram NLSF and MLE. This section, therefore, focuses on the bi-exponential case. Peak photon counts $A$ of 100, 200, 500, and 1{,}000 are considered alongside $\sigma_{IRF}$ values of 0.10, 0.15, 0.20, and 0.25 ns. These parameter ranges are representative of typical TCSPC-FLIM systems \cite{becker2005advanced}. The ground-truth (GT) lifetime parameters vary spatially, where $\tau_1$, $\tau_2$, and $\alpha_1$ take values of 0.3 ns, 2.0 ns, and 0.05 at the center, and 2.0 ns, 5.0 ns, and 0.95 at the edge, respectively. The results in all combinations of $A$ and $\sigma_{IRF}$ are shown in Fig.~\ref{fig:bi_A_sigma}. The reconstructed maps $\langle\tau\rangle$ in Fig. \ref{fig:bi_A_sigma} (a)-(d) recover the spatial structure of GT from all conditions tested for $M$ = 4, with performance improving as $A$ increases. Similar to existing studies using SSIM to evaluate reconstruction performance \cite{smith2019fast}, \cite{mannam2023improving}, we use it on the synthetic dataset evaluation. At $A=100$, reconstruction maps exhibit obvious noise. When $A=1{,}000$, the reconstructions show high fidelity to the GT in Fig.~\ref{fig:bi_A_sigma}(e), with smooth spatial transitions and suppressed noise. MAE decreases from 0.1071 ns to 0.034 ns as $A$ increases from 100 to 1,000, while SSIM improves from $\sim$0.17 to $\sim$0.57 and relative Accuracy from $\sim$0.78 to $\sim$0.93. The summary heatmaps in Fig.~\ref{fig:bi_A_sigma}(f) summarize the quantitative results of different $A$ and $\sigma_{IRF}$. Varying $\sigma_{IRF}$ from 0.10 to 0.25 ns has a slight effect on performance. That is because a wider IRF broadens the early-decay region, reducing the Fisher information concentrated near the rising edge with greater lifetime discrimination. Independent of the fitting-reconstruction algorithm, the Cramér-Rao bounds (CRB) are also considered under different settings of $A$ and $\sigma_{IRF}$, to show a lower bound for an unbiased estimator, as suggested in the Sketched-LiDAR paper \cite{spline2024}. In Fig.~\ref{fig:bi_A_sigma}(f), like MAE, the CRB decreases with $A$, and the effect of $\sigma_{IRF}$ is marginal. The Fisher-sketch reconstruction achieves $\sim$50\% efficiency relative to CRB across tested conditions, with RMSE approximately 2× the theoretical limit.

\begin{figure}[!t]
    \centering
    \includegraphics[width=1\linewidth]{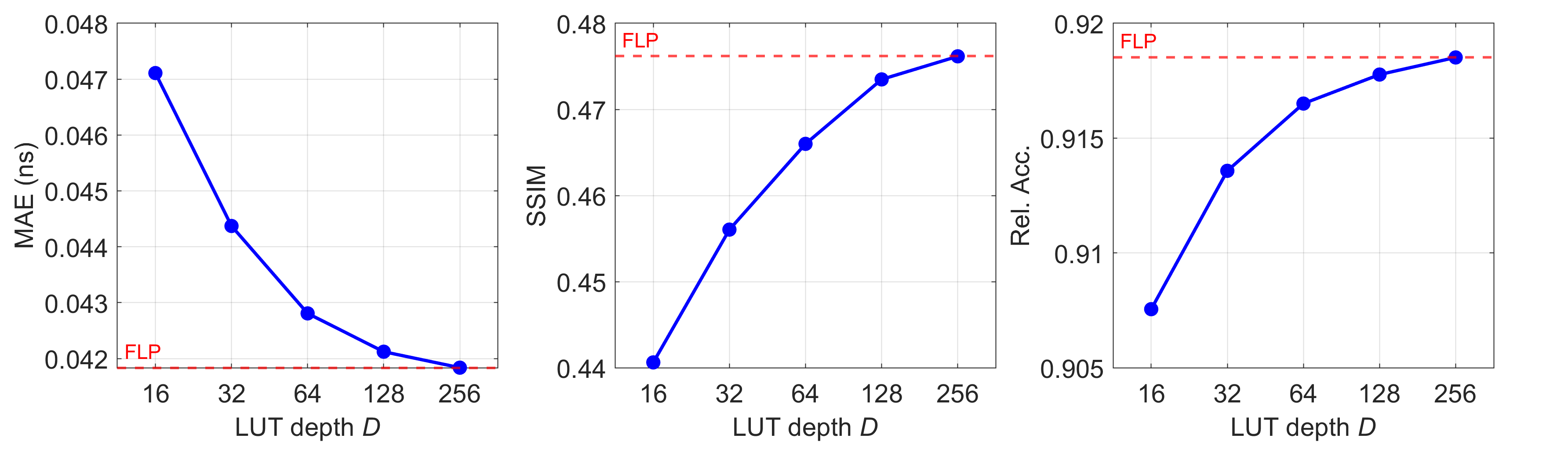}
    \caption{Reconstruction metrics of FXP LUT depth $D$, evaluated against the FLP baseline (red dashed).}
    \label{fig:lut_depth}
\end{figure}

\begin{figure*}[!t]
    \centering
    \includegraphics[width=1\linewidth]{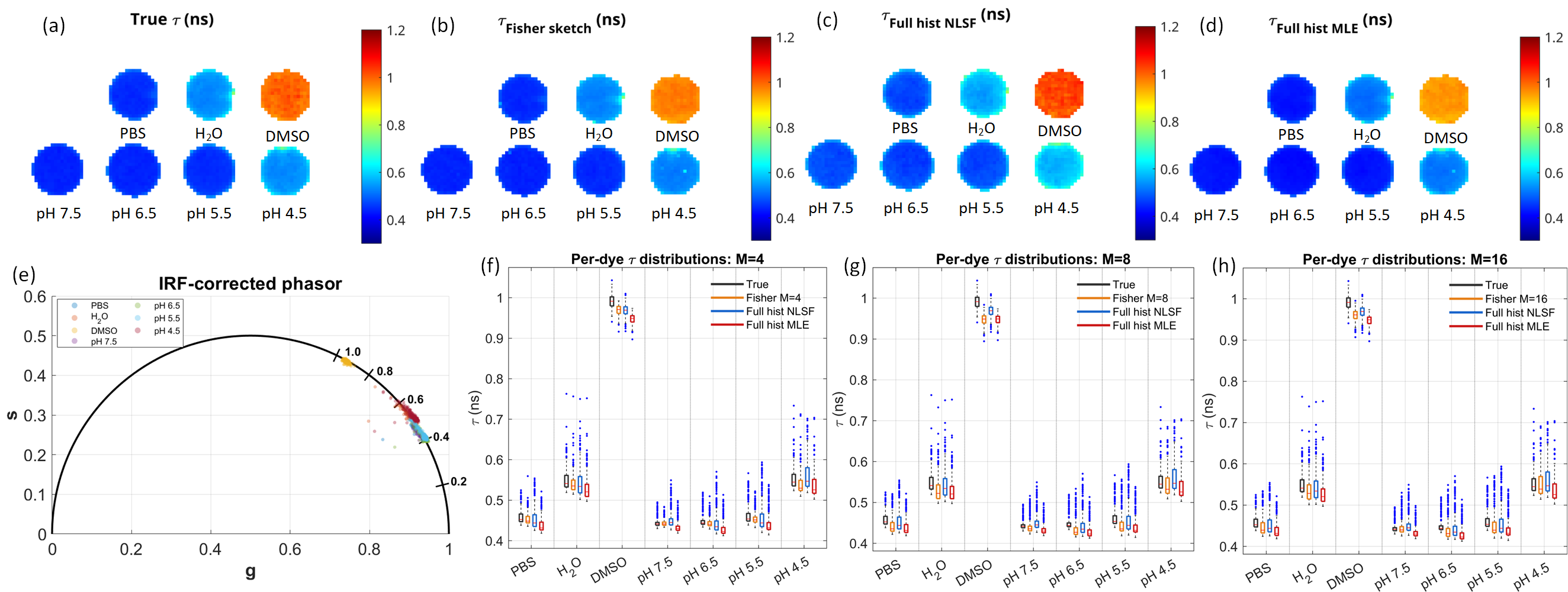}
    \caption{Experimental validation of Sketched-FLIM algorithm on mono-exponential FLIM data \cite{smith2022vitro}.
    (a) GT $\tau$ map. (b) Fisher-sketch reconstruction ($M=4$). 
    (c) Full-histogram NLSF. (d) Full-histogram  MLE.
    (e) IRF-corrected phasor plot.
    (f)-(h) Per-dye $\tau$ distributions across seven fluorescent dye solutions, retrieved from the Fisher-sketch method with $M=4, 8, 16$}.
    \label{fig:dye_mono}
\end{figure*}

To evaluate the effect of $M$ on reconstruction quality, $M$ varies from 4 to 20. For each condition, MAE, SSIM, and Relative Accuracy of $\langle\tau\rangle$ with a representative $\sigma_{IRF}=0.15$ ns, which is common in FLIM systems, to isolate the effect of $M$ and $A$. As shown in Fig.~\ref{fig:mean_tau_vs_M}, all three metrics improve consistently with increasing $M$, with the largest gain observed between $M=4$ and $M=8$. Beyond $M=8$, the improvement plateaus, meaning that a modest sketch dimension is sufficient to capture the lifetime-discriminating region of the decay. $A=1,000$ yields MAE below 0.03\,ns, SSIM above 0.60, and Relative Accuracy above 0.94, whereas at $A=100$ these degrade to approximately 0.09\,ns, 0.20, and 0.82. 

As different FLIM systems could operate with different time bins $N$, Fig.~\ref{fig:hist_length} evaluates the effect of $N$ on reconstruction quality, with $M=8$ fixed and metrics averaged over different $\sigma_{IRF}$. Unlike Fig.~\ref{fig:mean_tau_vs_M}, using peak photon count $A$, the total photon count is held constant. Increasing $N$ while keeping peak counts $A$ fixed increases the total number of detected photons, which is inconsistent with a fixed illumination budget. The budget photons are redistributed across finer time bins. Under this condition, three metrics exhibit a marginal degradation, remaining largely stable across the full range $N \in \{64, 128, 256, 512, 1,024\}$. Although finer time discretizations lead to a reduction in per-bin SNR, the sketch operator $\boldsymbol{W}$ in \ref{eq:sketch_matrix} proves robust to this effect, maintaining nearly consistent reconstruction fidelity. This robustness demonstrates that the proposed Sketched-FLIM framework generalizes reliably across FLIM systems with higher $N$. Therefore, the compression ratio can be achieved up to 128$\times$, or higher (256$\times$) when $M=4$, depending on the accuracy-complexity trade-off.

As the choice of $M=8$ induces a practical trade-off, we simulate its hardware implementation using a lookup table (LUT). Direct implementation of \ref{eq:linear_spline} on an FPGA requires division, which cannot be performed within a single clock cycle, introducing pipeline bubbles and a critical path. Instead, the spline basis values are precomputed offline using \ref{eq:linear_spline} and stored as a fixed-point (FXP) LUT in a 16-bit format with 8-bit integer and 8-bit fractional parts (16-bit, 8I+8F), a common format in digital logic design. The LUT depth $D$ balances hardware resource consumption and approximation precision. Therefore, we evaluate $D \in \{16, 32, 64, 128, 256\}$ using the same synthetic data and metrics as in Fig.~\ref{fig:mean_tau_vs_M}. In Fig.~\ref{fig:lut_depth}, the three metrics converge monotonically to the floating-point (FLP) baseline as $D$ increases, with $D=128$ achieving near-identical performance at substantially reduced hardware cost. $D=256$ reaches the same accuracy as the FLP baseline because $N=256$.

\begin{figure*}[!t]
    \centering
    \includegraphics[width=1\linewidth]{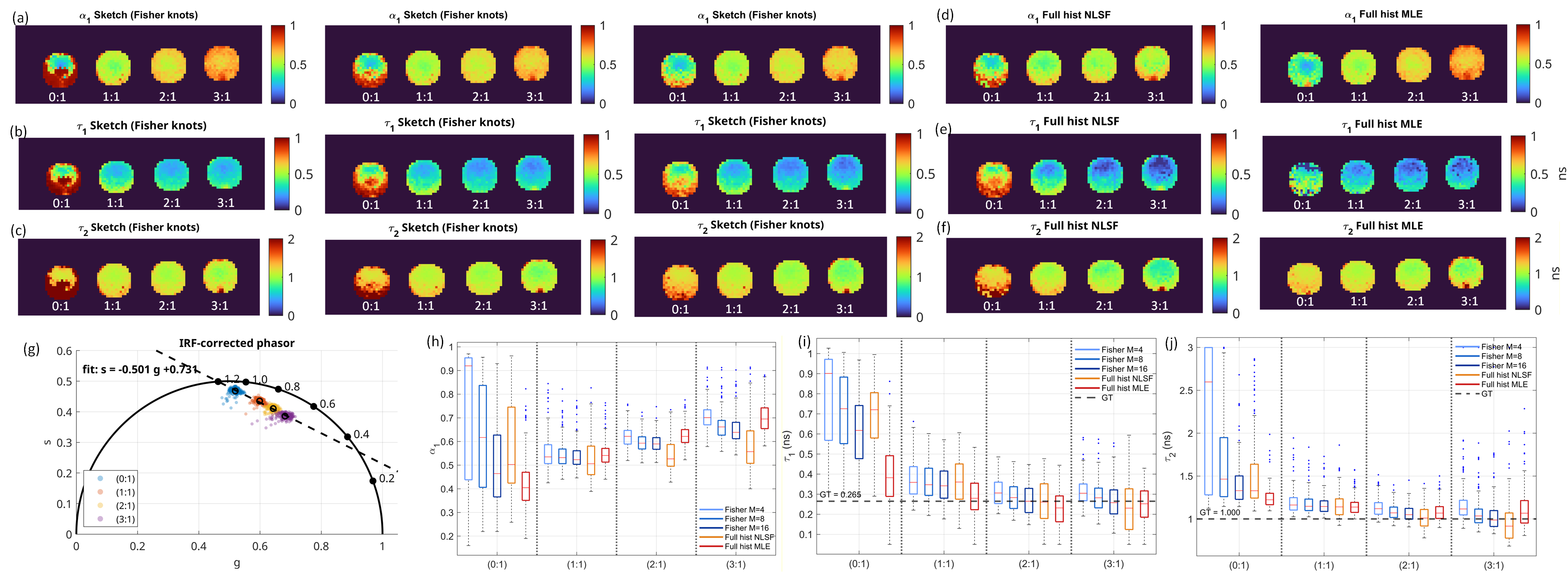}
    \caption{Experimental validation of Sketched-FLIM algorithm on bi-exponential FLIM data \cite{smith2022vitro}.
    (a)--(c)~Spatial maps of $\alpha_1$, $\tau_1$, and $\tau_2$ for Fisher sketch at $M=4$, 8, and 16 (columns), and (d)--(f) the corresponding NLSF and MLE baselines. (g)~IRF-corrected Phasor plot.
    (h)--(j)~Per-sample distributions of $\alpha_1$, $\tau_1$, and $\tau_2$ 
    across four acceptor-to-donor ratios.}
    \label{fig:dye_bi}
\end{figure*}

\section{Real Datasets Evaluation}
\label{sec:real_data_evaluation}
Real datasets are adopted from a publicly available \textit{in vitro} FLIM dataset~\cite{smith2022vitro}, licensed under CC~BY~4.0. The raw data are histograms with 251 time bins and a bin width of 39.84 ps. As in the synthetic dataset evaluation, timestamps are converted from the histogram for our Fisher-sketch processing. The dataset includes measurements of seven dye solutions (PBS, H$_2$O, DMSO, and pH 7.5--4.5) for mono-exponential validation, and an $\alpha_1$ plate with four acceptor-to-donor ratios (0:1 to 3:1) for bi-exponential validation. The proposed Sketched-FLIM is evaluated by comparing  to NLSF and MLE baselines. Three sketch dimensions $M \in \{4, 8, 16\}$ are assessed by investigating the effect of $M$ on real data. It should be noted that the original study ~\cite{smith2022vitro} determines $\tau_1$ and $\tau_2$ from independent measurements and treats them as fixed GT values, fitting only the FRET intensity fraction per pixel. In contrast, Sketched-FLIM reconstructs all three bi-exponential parameters $\tau_1$, $\tau_2$, and $\alpha_1$ simultaneously per pixel, without GT $\tau_1$ or $\tau_2$.

Results for the mono-exponential case are presented in Fig.~\ref{fig:dye_mono}. The reconstructed $\tau$ maps in Fig.~\ref{fig:dye_mono}(b) closely reproduce the GT spatial structure in Fig.~\ref{fig:dye_mono}(a), with dye-specific lifetime contrast preserved across all beads. The IRF-corrected Phasor plot in Fig.~\ref{fig:dye_mono}(e) is computed from timestamps directly without building a histogram, as proved in Appendix \ref{sec:phasor_timestamp}. The phasor plot aligns with the one computed from full histograms in \cite{smith2022vitro}. As per-dye's $\tau$ distributions shown in Fig.~\ref{fig:dye_mono}(f), Sketched-FLIM with $M=4$ produces median estimates aligning with both NLSF and MLE in all conditions. As $M$ increases from 4 to 16, the interquartile range narrows progressively, indicating improved precision, while the median remains stable. This is consistent with the synthetic results in Section~\ref{sec:synth_results}, where higher $M$ yields better reconstruction fidelity with marginal improvement when $M>8$. Results for the bi-exponential $\alpha_1$ case are shown in Fig.~\ref{fig:dye_bi}. Reconstruction of $\alpha_1$, $\tau_1$, and $\tau_2$ in Fig.~\ref{fig:dye_bi}(a)--(f) demonstrates that Sketched-FLIM recovers the expected $\alpha_1$-inducing parameter structure across all four acceptor-to-donor conditions. Similar to the phasor plot for mono-exponential data, the phasor plot in Fig.~\ref{fig:dye_bi}(g) is also from timestamps. Four conditions of mixture cluster along a linear trajectory, as expected for a two-component mixture, with the linear fit agreeing with the phasor in \cite{smith2022vitro} from full histograms. The per-sample distributions in Fig.~\ref{fig:dye_bi}(h)--(j) show that the Fisher-sketch with $M=8$ and $M=16$ produces $\alpha_1$, $\tau_1$, and $\tau_2$ estimations that are close to NLSF and MLE, while $M=4$ exhibits slightly wider distributions, particularly for $\tau_1$ because $\alpha_1 \approx 0$ for 0:1 condition and the decay is dominated by the long-lifetime component $\tau_2$, making $\tau_1$ and $\alpha_1$ poorly identifiable from the signal. As the ratio increases toward (3:1), the short-lifetime component $\tau_1$ contributes more significantly to the decay shape, improving the identifiability of all three parameters, indicated by increasing accuracy in Fig.\ref{fig:dye_bi}(h)---(j). Poisson MLE achieves the tightest distributions. Notably, the Fisher-sketch achieves median estimates comparable to full-histogram NLSF while exhibiting smaller interquartile ranges. As mentioned, Fisher-sketch-based knot allocation concentrates on high-SNR regions and downweights bins with sparse photon counts, whereas full-histogram NLSF weights each bin equally. This Fisher-optimal knot placement suppresses the influence of noisy bins, alleviating the fitting instability. Overall, the bi-exponential results confirm that Sketched-FLIM generalizes effectively to real multi-component FLIM data, achieving competitive accuracy with a substantially reduced data volume (62.75$\times$ in current datasets) compared to conventional full-histogram methods.

\section{Conclusion}
\label{sec:conclusion}
We introduce Sketched-FLIM, which processes photon timestamps directly and reconstructs fluorescence-lifetime parameters for both mono- and bi-exponential decays. We further establish a connection between the Fourier sketching scheme and the phasor method in the frequency domain, revealing the interpretability of the Fourier sketch and motivating a progression from linear, uniform spline-knot allocation to Fisher-information-guided, non-uniform knot placement. Extensive evaluation on synthetic datasets across multiple dimensions, including histogram bin count $N$, IRF width $\sigma_{IRF}$, photon count, and the number of spline channels $M$, demonstrates that Sketched-FLIM achieves performance comparable to full-histogram fitting solutions, without requiring histogram storage. Evaluation on experimental datasets further validates the consistency between full-histogram-based fitting and Sketched-FLIM. Finally, an FXP LUT-based simulation of the hardware implementation paves the way for firmware integration in future work.


\appendix

\section{The phasor Representation as a Special Case of the Fourier Sketch}
\label{app:phasor}

\subsection{Equivalence Between \texorpdfstring{$M=1$}{M=1} Fourier Sketch and Phasor}
\label{app:phasor_A}

Let $h(t_k)$ denote the fluorescence decay histogram sampled over $N$
time bins within the acquisition.
The general Fourier sketch projects the histogram onto complex
exponential bases at harmonic index $m$:
\begin{equation}
    z_m = \sum_{k=0}^{N-1} h(t_k)\,
          \exp\!\left(-j\frac{2\pi m}{T}t_k\right).
    \label{eq:fourier_sketch_general}
\end{equation}
When only the first harmonic is retained ($M=1$), this becomes
\begin{equation}
    z_1 = \sum_{k=0}^{N-1} h(t_k)\,
          \exp\!\left(-j\frac{2\pi}{T}t_k\right).
    \label{eq:fourier_sketch_m1}
\end{equation}
Expanding the complex exponential gives
\begin{equation}
    z_1 = \sum_{k=0}^{N-1} h(t_k)
          \left[
            \cos\!\left(\frac{2\pi}{T}t_k\right)
            - j\sin\!\left(\frac{2\pi}{T}t_k\right)
          \right].
    \label{eq:fourier_expanded}
\end{equation}
Separating real and imaginary parts obtains
\begin{align}
    \mathrm{Re}(z_1) &= \sum_{k=0}^{N-1}
                         h(t_k)\cos\!\left(\frac{2\pi}{T}t_k\right),
    \label{eq:re_z1} \\
    \mathrm{Im}(z_1) &= -\sum_{k=0}^{N-1}
                          h(t_k)\sin\!\left(\frac{2\pi}{T}t_k\right).
    \label{eq:im_z1}
\end{align}
In FLIM, the phasor coordinates are defined as the normalized real and imaginary parts of the first Fourier harmonic~\cite{digman2008phasor}:
\begin{align}
    g &= \frac{\displaystyle\sum_{k} h(t_k)
               \cos\!\left(\tfrac{2\pi}{T}t_k\right)}
              {\displaystyle\sum_{k} h(t_k)},
    \label{eq:phasor_g} \\[4pt]
    s &= \frac{\displaystyle\sum_{k} h(t_k)
               \sin\!\left(\tfrac{2\pi}{T}t_k\right)}
              {\displaystyle\sum_{k} h(t_k)}.
    \label{eq:phasor_s}
\end{align}
Comparing \eqref{eq:re_z1}--\eqref{eq:im_z1} with
\eqref{eq:phasor_g}--\eqref{eq:phasor_s} shows that
$(g, s) \propto \bigl(\mathrm{Re}(z_1),\,-\mathrm{Im}(z_1)\bigr)$,
i.e.\ the phasor is the normalized $M=1$ Fourier sketch.
Higher-order sketches ($M>1$) extend the phasor by considering additional harmonics of the decay signal, capturing more lifetime information. For a single-exponential fluorescence decay $I(t) = I_0\,e^{-t/\tau}$, the phasor coordinates are obtained by evaluating the normalized Fourier transform at angular frequency
$\omega = 2\pi/T$:
\begin{align}
    g(\tau) &= \frac{\int_0^{\infty} e^{-t/\tau}\cos(\omega t)\,dt}
                    {\int_0^{\infty} e^{-t/\tau}\,dt}
             = \frac{1}{1+(\omega\tau)^2},
    \label{eq:g_tau} \\[4pt]
    s(\tau) &= \frac{\int_0^{\infty} e^{-t/\tau}\sin(\omega t)\,dt}
                    {\int_0^{\infty} e^{-t/\tau}\,dt}
             = \frac{\omega\tau}{1+(\omega\tau)^2}.
    \label{eq:s_tau}
\end{align}
Setting $u = \omega\tau \ge 0$ and computing
$\bigl(g - \tfrac{1}{2}\bigr)^2 + s^2$:
\begin{align}
    \left(\frac{1}{1+u^2} - \frac{1}{2}\right)^2
    + \left(\frac{u}{1+u^2}\right)^2
    = \frac{(1+u^2)^2}{4(1+u^2)^2}
     = \frac{1}{4}.
    \label{eq:semicircle_proof}
\end{align}
Therefore, for all $\tau > 0$, the phasor satisfies
$(g - \tfrac{1}{2})^2 + s^2 = \tfrac{1}{4}$,
confirming that all single-exponential phasor points lie on
the universal semicircle of radius $\tfrac{1}{2}$ centered
at $(\tfrac{1}{2}, 0)$. Multi-exponential decays correspond to combinations of single-exponential phasors and therefore lie strictly inside the semicircle, providing the geometric basis for phasor-based lifetime unmixing.

\subsection{Direct Accumulation from Timestamps}
\label{sec:phasor_timestamp}
In TCSPC, photon arrivals are recorded as timestamps
$\{t_i\}_{i=1}^{P}$.
The Fourier sketch coefficient at harmonic $m$ is accumulated
directly from timestamps
\begin{equation}
    z_m = \sum_{i=1}^{P}
          \exp\!\left(j\frac{2\pi m}{T}t_i\right).
    \label{eq:sketch_timestamps}
\end{equation}
Substituting the histogram $h(t_k) = \sum_i 1(t_i \in [t_k, t_{k+1}))$
into \eqref{eq:fourier_sketch_general} recovers \eqref{eq:sketch_timestamps}. Each arriving photon contributes one complex multiplication to the running phasor sum, which can be implemented in firmware using LUTs.

\section*{Acknowledgements}
The authors acknowledge Smith~\textit{et~al.}\ for making the 
\textit{in vitro} FLIM dataset publicly available~\cite{smith2022vitro}, which was used for experimental validation in this work.
\bibliographystyle{IEEEtran}
\bibliography{reference}

@book{lakowicz2006principles,
  title={Principles of fluorescence spectroscopy},
  author={Lakowicz, Joseph R},
  year={2006},
  publisher={Springer}
}

@article{torrado2024fluorescence,
  title={Fluorescence lifetime imaging microscopy},
  author={Torrado, Belen and Pannunzio, Bruno and Malacrida, Leonel and Digman, Michelle A},
  journal={Nature Reviews Methods Primers},
  volume={4},
  number={1},
  pages={80},
  year={2024},
  publisher={Nature Publishing Group UK London}
}

@book{becker2005advanced,
  title={Advanced time-correlated single photon counting techniques},
  author={Becker, Wolfgang},
  year={2005},
  publisher={Springer}
}

@article{henderson2019192,
  title={A 192$\times$128 time correlated SPAD image sensor in 40-nm CMOS technology},
  author={Henderson, Robert K and Johnston, Nick and Della Rocca, Francescopaolo Mattioli and Chen, Haochang and Li, David Day-Uei and Hungerford, Graham and Hirsch, Richard and Mcloskey, David and Yip, Philip and Birch, David JS},
  journal={IEEE Journal of Solid-State Circuits},
  volume={54},
  number={7},
  pages={1907--1916},
  year={2019},
  publisher={IEEE}
}

@article{heliot2021simple,
  title={Simple phasor-based deep neural network for fluorescence lifetime imaging microscopy},
  author={H{\'e}liot, Laurent and Leray, Aymeric},
  journal={Scientific reports},
  volume={11},
  number={1},
  pages={23858},
  year={2021},
  publisher={Nature Publishing Group UK London}
}

@article{sahoo2011forster,
  title={F{\"o}rster resonance energy transfer--A spectroscopic nanoruler: Principle and applications},
  author={Sahoo, Harekrushna},
  journal={Journal of Photochemistry and Photobiology C: Photochemistry Reviews},
  volume={12},
  number={1},
  pages={20--30},
  year={2011},
  publisher={Elsevier}
}

@article{shirshin2022label,
  title={Label-free sensing of cells with fluorescence lifetime imaging: The quest for metabolic heterogeneity},
  author={Shirshin, Evgeny A and Shirmanova, Marina V and Gayer, Alexey V and Lukina, Maria M and Nikonova, Elena E and Yakimov, Boris P and Budylin, Gleb S and Dudenkova, Varvara V and Ignatova, Nadezhda I and Komarov, Dmitry V and others},
  journal={Proceedings of the National Academy of Sciences},
  volume={119},
  number={9},
  pages={e2118241119},
  year={2022},
  publisher={National Academy of Sciences}
}

@article{smith2022vitro,
  title={In vitro and in vivo NIR fluorescence lifetime imaging with a time-gated SPAD camera},
  author={Smith, Jason T and Rudkouskaya, Alena and Gao, Shan and Gupta, Juhi M and Ulku, Arin and Bruschini, Claudio and Charbon, Edoardo and Weiss, Shimon and Barroso, Margarida and Intes, Xavier and others},
  journal={Optica},
  volume={9},
  number={5},
  pages={532--544},
  year={2022},
  publisher={Optica Publishing Group}
}

@article{zang2022fast,
  title={Fast analysis of time-domain fluorescence lifetime imaging via extreme learning machine},
  author={Zang, Zhenya and Xiao, Dong and Wang, Quan and Li, Zinuo and Xie, Wujun and Chen, Yu and Li, David Day Uei},
  journal={Sensors},
  volume={22},
  number={10},
  pages={3758},
  year={2022},
  publisher={MDPI}
}

@article{wang2024deep,
  title={Deep learning-based virtual H\&E staining from label-free autofluorescence lifetime images},
  author={Wang, Qiang and Akram, Ahsan R and Dorward, David A and Talas, Sophie and Monks, Basil and Thum, Chee and Hopgood, James R and Javidi, Malihe and Vallejo, Marta},
  journal={npj Imaging},
  volume={2},
  number={1},
  pages={17},
  year={2024},
  publisher={Nature Publishing Group UK London}
}

@article{lin2024coupling,
  title={Coupling a recurrent neural network to SPAD TCSPC systems for real-time fluorescence lifetime imaging},
  author={Lin, Yang and Mos, Paul and Ardelean, Andrei and Bruschini, Claudio and Charbon, Edoardo},
  journal={Scientific Reports},
  volume={14},
  number={1},
  pages={3286},
  year={2024},
  publisher={Nature Publishing Group UK London}
}

@inproceedings{lin2024spiking,
  title={Spiking neural networks for active time-resolved SPAD imaging},
  author={Lin, Yang and Charbon, Edoardo},
  booktitle={Proceedings of the IEEE/CVF Winter Conference on Applications of Computer Vision},
  pages={8147--8156},
  year={2024}
}

@article{erdogan2019cmos,
  title={A CMOS SPAD line sensor with per-pixel histogramming TDC for time-resolved multispectral imaging},
  author={Erdogan, Ahmet T and Walker, Richard and Finlayson, Neil and Krstaji{\'c}, Nikola and Williams, Gareth and Girkin, John and Henderson, Robert},
  journal={IEEE Journal of Solid-State Circuits},
  volume={54},
  number={6},
  pages={1705--1719},
  year={2019},
  publisher={IEEE}
}

@ARTICLE{spline2024,
  author={Sheehan, Michael P. and Tachella, Julián and Davies, Mike E.},
  journal={IEEE Transactions on Computational Imaging}, 
  title={Spline Sketches: An Efficient Approach for Photon Counting Lidar}, 
  year={2024},
  volume={10},
  number={},
  pages={863-875},
  doi={10.1109/TCI.2024.3404652}}

@article{sheehan2021sketching,
  title={A sketching framework for reduced data transfer in photon counting lidar},
  author={Sheehan, Michael P and Tachella, Juli{\'a}n and Davies, Mike E},
  journal={IEEE Transactions on Computational Imaging},
  volume={7},
  pages={989--1004},
  year={2021},
  publisher={IEEE}
}

@article{zang2026fpga,
  title={FPGA Implementation of Sketched LiDAR for a 192 x 128 SPAD Image Sensor},
  author={Zang, Zhenya and Davies, Mike and Gyongy, Istvan},
  journal={arXiv preprint arXiv:2602.10837},
  year={2026}
}

@book{cramer1999,
  title={Mathematical methods of statistics},
  author={Cram{\'e}r, Harald},
  volume={9},
  year={1999},
  publisher={Princeton University Press}
}

@article{rao1945information,
  title={Information and the accuracy attainable in the estimation of statistical parameters},
  author={Rao, C Radhakrishna and others},
  journal={Bull. Calcutta Math. Soc},
  volume={37},
  number={3},
  pages={81--91},
  year={1945}
}

@article{digman2008phasor,
  title={The phasor approach to fluorescence lifetime imaging analysis},
  author={Digman, Michelle A and Caiolfa, Valeria R and Zamai, Moreno and Gratton, Enrico},
  journal={Biophysical journal},
  volume={94},
  number={2},
  pages={L14--L16},
  year={2008},
  publisher={Elsevier}
}

@article{li2020investigations,
  title={Investigations on average fluorescence lifetimes for visualizing multi-exponential decays},
  author={Li, Yahui and Natakorn, Sapermsap and Chen, Yu and Safar, Mohammed and Cunningham, Margaret and Tian, Jinshou and Li, David Day-Uei},
  journal={Frontiers in physics},
  volume={8},
  pages={576862},
  year={2020},
  publisher={Frontiers Media SA}
}

@article{zang2023compact,
  title={Compact and robust deep learning architecture for fluorescence lifetime imaging and FPGA implementation},
  author={Zang, Zhenya and Xiao, Dong and Wang, Quan and Jiao, Ziao and Chen, Yu and Li, David Day Uei},
  journal={Methods and applications in fluorescence},
  volume={11},
  number={2},
  pages={025002},
  year={2023},
  publisher={IOP Publishing}
}

@article{malacrida2021phasor,
  title={The phasor plot: a universal circle to advance fluorescence lifetime analysis and interpretation},
  author={Malacrida, Leonel and Ranjit, Suman and Jameson, David M and Gratton, Enrico},
  journal={Annual review of biophysics},
  volume={50},
  number={1},
  pages={575--593},
  year={2021},
  publisher={Annual Reviews}
}

@article{smith2019fast,
  title={Fast fit-free analysis of fluorescence lifetime imaging via deep learning},
  author={Smith, Jason T and Yao, Ruoyang and Sinsuebphon, Nattawut and Rudkouskaya, Alena and Un, Nathan and Mazurkiewicz, Joseph and Barroso, Margarida and Yan, Pingkun and Intes, Xavier},
  journal={Proceedings of the national academy of sciences},
  volume={116},
  number={48},
  pages={24019--24030},
  year={2019},
  publisher={National Academy of Sciences}
}

@article{mannam2023improving,
  title={Improving fluorescence lifetime imaging microscopy phasor accuracy using convolutional neural networks},
  author={Mannam, Varun and P. Brandt, Jacob and Smith, Cody J and Yuan, Xiaotong and Howard, Scott},
  journal={Frontiers in Bioinformatics},
  volume={3},
  pages={1335413},
  year={2023},
  publisher={Frontiers Media SA}
}

\end{document}